# The integral theorem of generalized virial in the relativistic uniform model

**Sergey G. Fedosin**

PO box 614088, Sviazeva str. 22-79, Perm, Perm Krai, Russia

e-mail fedosin@hotmail.com

In the relativistic uniform model for continuous medium the integral theorem of generalized virial is derived, in which generalized momenta are used as particles' momenta. This allows us to find exact formulas for the radial component of the velocity of typical particles of the system and for their root-mean-square speed, without using the notion of temperature. The relation between the theorem and the cosmological constant, characterizing the physical system under consideration, is shown. The difference is explained between the kinetic energy and the energy of motion, the value of which is equal to half the sum of the Lagrangian and the Hamiltonian. This difference is due to the fact that the proper fields of each particle have mass-energy, which makes an additional contribution into the kinetic energy. As a result, the total energy of motion of particles and fields is obtained.

**Keywords:** generalized virial theorem; relativistic uniform model; cosmological constant; energy of motion; kinetic energy.

## 1. Introduction

In theoretical physics, the virial theorem is a relation between the kinetic energy and other types of energy in a system of particles and fields. There are various modifications of the theorem, in particular in classical mechanics [1], in analytical Lagrangian mechanics [2] and in quantum mechanics [3]. As a rule, vector notation of the theorem is used, but tensor variants are also possible [4].

Earlier we studied application of the virial theorem in a relativistic uniform system consisting of particles held by their proper fields [5]. Thereby we obtained the difference from the classical case, reaching the value of 20%. The reason for this situation was inequality to zero of the convective part of the time derivative of the system's virial function averaged over time, calculated by us.

Now we would like to consider the theorem of generalized virial (rather than ordinary virial) in the relativistic uniform model and to apply the possibilities it provides. In particular, we will be able to find exact formulas for the radial component of the velocity of the particles inside the system and for the root-mean-square speed, as well as understand the difference between the kinetic energy and the energy of the particles' motion associated with their generalized momenta.

## 2. The generalized virial theorem

We will define the generalized virial function as follows:

$$G_V = \sum_{n=1}^{N} \left( \mathbf{P}_n \cdot \mathbf{r}_n \right), \tag{1}$$

where $\mathbf{P}_n$ is the generalized three-momentum of an arbitrary particle of the system; $\mathbf{r}_n$ is the three-vector of location of the particle with the number $n$, this vector is included in the number of possible generalized coordinates; $N$ specifies the number of particles in the system.

It should be noted that all the fields associated with the given particle and having influence on it make contribution to the generalized momentum $\mathbf{P}_n$ of the particle. We will use the generalized momenta in (1) because in a closed system the sum of such particles' momenta is conserved [6]. In contrast to this, in the usual formulation of the virial theorem, instead of $\mathbf{P}_n$ in (1) there is the momentum of the particle with the number $n$, found through the mass and velocity of the particle.

In the case under consideration, the generalized three-velocity of the particle is given by the expression $\mathbf{v}_n = \dfrac{d\mathbf{r}_n}{dt}$. For continuously distributed systems, it can be seen from (1) that a change in the generalized virial function with respect to a certain center can be associated with both a change in the particles' momenta and a certain change in the system's shape.

Let us take the time derivative of the virial function (1):

$$\frac{dG_V}{dt} = \sum_{n=1}^{N} \left( \frac{d\mathbf{P}_n}{dt} \cdot \mathbf{r}_n \right) + \sum_{n=1}^{N} \left( \mathbf{P}_n \cdot \mathbf{v}_n \right). \tag{2}$$

According to the standard procedure, it is also necessary to perform time-averaging of all the terms in equation (2) over a sufficiently large period of time. In many practical cases, the left-hand side of (2) tends to or is close to zero, which leads to the relationship between the two terms on the right-hand side. By the order of magnitude the sum $\sum_{n=1}^{N}\left(\dfrac{d\mathbf{P}_n}{dt}\cdot\mathbf{r}_n\right)=\sum_{n=1}^{N}(\mathbf{F}_n\cdot\mathbf{r}_n)$, where the three-vector $\mathbf{F}_n$ is the generalized force, is equal to the potential energy $W$ of the particles' interaction in the case of potential forces, and the sum $\sum_{n=1}^{N}(\mathbf{P}_n\cdot\mathbf{v}_n)$ is approximately equal to the doubled kinetic energy $E_k$. Under these assumptions, the classical virial theorem follows from (2):

$$E_k \approx -0.5W. \tag{3}$$

### 3. The relativistic uniform system

Relativistic uniformity implies that the invariant density or the mass density (charge density) in the reference frames associated with the particles is constant for all the particles.

Suppose there is a spherical system of such closely interacting particles, which are bound to each other by the gravitational and electromagnetic fields. We will also use the concept of the vector pressure field, as well as the concept of the vector acceleration field, in which the role of the stress-energy tensor of the matter is played by the stress-energy tensor of the acceleration field [7, 8]. All the four fields are vector fields, they can be formed by the same pattern, they have the proper four-potentials, and therefore the corresponding scalar and vector potentials.

For the case when the particles interact so closely with each other that they practically merge and form continuously distributed matter, the so-called typical particles are taken as representative units of the matter. The equations of motion are applied to typical particles, and all the physical quantities are also written as applied to typical particles. It is assumed that typical particles on the average characterize the matter in all respects. For typical particles of the continuously distributed matter, it is convenient to rewrite (1) in terms of the integral over the volume, using the results in [7]:

$$\mathbf{P}_n = \dfrac{1}{c}\int^n \left(\rho_0 \mathbf{U}_n + \rho_0 \mathbf{D}_n + \rho_{0q}\mathbf{A}_n + \rho_0 \mathbf{\Pi}_n\right) u^0 \sqrt{-g}\, dx^1 dx^2 dx^3, \tag{4}$$

$$G_V = \dfrac{1}{c}\sum_{n=1}^{N}\mathbf{r}_n \cdot \int^n \left(\rho_0 \mathbf{U}_n + \rho_0 \mathbf{D}_n + \rho_{0q}\mathbf{A}_n + \rho_0 \mathbf{\Pi}_n\right) u^0 \sqrt{-g}\, dx^1 dx^2 dx^3,$$

where $c$ is the speed of light; $\rho_0$ is the invariant mass density; $\rho_{0q}$ is the invariant charge density; $\mathbf{U}$, $\mathbf{D}$, $\mathbf{A}$ and $\mathbf{\Pi}$ represent the vector potentials of the acceleration field, gravitational field, electromagnetic field and pressure field, respectively; $u^0$ is the time component of the particle's four-velocity.

In (4), the symbol $\int^n$ denotes the integral over the volume of one moving typical particle, and the symbol $\sum_{n=1}^{N}$ implies summation over all particles.

Next we will consider the weak field approximation, when the spacetime curvature can be neglected and the situation can be considered in the Minkowski flat spacetime in the framework of the special theory of relativity. In this case, the determinant of the metric tensor equals $g = -1$, and the element of the covariant volume $\sqrt{-g}\, dx^1 dx^2 dx^3$ is replaced by the element of the ordinary three-dimensional volume $dV$.

Consequently, the quantity $u^0 = c\gamma'$ in (4) will be the time component of the average four-velocity of the particles located at the current radius $r$, while the Lorentz factor $\gamma'$ of the particles inside the sphere according to [9] turns out to be a function of the current radius:

$$\gamma' = \frac{c\gamma_c}{r\sqrt{4\pi\eta\rho_0}} \sin\left(\frac{r}{c}\sqrt{4\pi\eta\rho_0}\right) \approx \gamma_c - \frac{2\pi\eta\rho_0 r^2 \gamma_c}{3c^2}. \tag{5}$$

Here $\eta$ is the acceleration field coefficient, $\gamma_c = \dfrac{1}{\sqrt{1 - v_c^2/c^2}}$ is the Lorentz factor for the speed $v_c$ of the particles at the center of the sphere, and, due to of the smallness of the argument, the sine can be expanded up to the second-order terms.

In order to simplify further calculations, we will assume that the particles in the system under consideration move randomly without general rotation and directed fluxes of matter. In this case, the global vector potentials of each field vanish, since the vector sum of the potentials at an arbitrary point tends to zero because of the different directions of the vector potentials of individual particles. For each of the particles only their own vector potentials are left, arising from the motion of their proper internal fields.

Therefore, in (4) $\mathbf{U}$, $\mathbf{D}$, $\mathbf{A}$ and $\mathbf{\Pi}$ should be replaced by the small proper vector potentials $\mathbf{U}_p$, $\mathbf{D}_p$, $\mathbf{A}_p$ and $\mathbf{\Pi}_p$, which are inversely proportional to the square of the speed of light and directly proportional to the particles' velocities $\mathbf{v}'$ and the scalar potentials $\vartheta_p$, $\psi_p$, $\varphi_p$ and $\wp_p$ of the proper internal fields of the particles. Proceeding similarly to [5], in the approximation of rectilinear motion of the particles without acceleration, which is typical of the special theory of relativity, we have the following:

$$\mathbf{U}_p = \frac{\vartheta_p \mathbf{v}'}{c^2} = \frac{\vartheta_{0p} \gamma' \mathbf{v}'}{c^2}, \qquad \mathbf{D}_p = \frac{\psi_p \mathbf{v}'}{c^2} = \frac{\psi_{0p} \gamma' \mathbf{v}'}{c^2},$$

$$\mathbf{A}_p = \frac{\varphi_p \mathbf{v}'}{c^2} = \frac{\varphi_{0p} \gamma' \mathbf{v}'}{c^2}, \qquad \mathbf{\Pi}_p = \frac{\wp_p \mathbf{v}'}{c^2} = \frac{\wp_{0p} \gamma' \mathbf{v}'}{c^2}. \qquad (6)$$

The scalar potentials $\vartheta_{0p}$, $\psi_{0p}$, $\varphi_{0p}$ and $\wp_{0p}$ of the acceleration field, gravitational and electromagnetic fields and pressure field, respectively, are the potentials in the center-of-momentum frame of each particle. We can assume that if the particles contain the randomly moving matter and represent relativistic uniform systems, they do not have the internal global vector potentials. But for an observer, who is stationary relative to the sphere, the particles are moving, and this observer notes inside the particles both the scalar potentials $\vartheta_p$, $\psi_p$, $\varphi_p$ and $\wp_p$, and the vector potentials (6). Substitution of (6) into (4) within the framework of the special theory of relativity gives the following:

$$G_V = \frac{1}{c^2} \sum_{n=1}^{N} \gamma_n'^2 \mathbf{r}_n \cdot \mathbf{v}_n' \int^n \left( \rho_0 \vartheta_{0p} + \rho_0 \psi_{0p} + \rho_{0q} \varphi_{0p} + \rho_0 \wp_{0p} \right) dV. \qquad (7)$$

As was shown in [9, 10], in the first approximation the scalar potentials of the fields inside the relativistic uniform system depend only on the square of the radius. In application to an individual particle, they can be expressed as follows:

$$\vartheta_{0p} = \gamma_p c^2 \approx c^2 \gamma_{cp} - \frac{2\pi \eta \rho_{0s} r^2 \gamma_{cp}}{3}, \qquad \psi_{0p} \approx \frac{2\pi G \gamma_{cp} \rho_{0s}(r^2 - 3a_p^2)}{3},$$

$$\varphi_{0p} \approx \frac{\gamma_{cp} \rho_{0qs}(3a_p^2 - r^2)}{6\varepsilon_0}, \qquad \wp_{0p} \approx \wp_{cp} - \frac{2\pi \sigma \rho_{0s} r^2 \gamma_{cp}}{3}, \qquad (8)$$

where $\gamma_p$ is the Lorentz factor of the subparticles' motion inside the particle, written similarly to (5); $\gamma_{cp}$ is the Lorentz factor at the center of the particle; $\rho_{0s}$ is the invariant mass density of the subparticles; $r$ is the current radius inside the particle; $G$ is the gravitational constant; $a_p$ is the radius of the particle; $\varepsilon_0$ is the electrical constant; $\rho_{0qs}$ is the invariant charge density of the subparticles; $\wp_{cp}$ is the scalar potential of the pressure field at the center of the particle; $\sigma$ is the pressure field coefficient.

The volume element $dV$ in (7) is an element of the volume of a fixed sphere. In the first approximation, we can assume that $dV$ is also an element of the moving volume $dV_p$ of the particle, then the sum of such volumes over all the particles must give the volume of the sphere. Therefore, the integral in (7) can be considered as the integral over the moving volume of the particle with the number $n$.

Let us assume that the moving particle at the initial time point crosses the origin of the fixed reference frame, moving at a constant speed $v'$ along the axis $OX$. Proceeding further according to [11] and Appendix A in [10], we will express the coordinates inside the particle from the perspective of the fixed reference frame in terms of the spherical coordinates $\rho, \theta, \varphi$ in the center-of-momentum frame of the particle:

$$x = \sqrt{1 - v'^2/c^2}\, \rho \cos\theta + v't, \qquad y = \rho \sin\theta \cos\varphi, \qquad z = \rho \sin\theta \sin\varphi. \tag{9}$$

The scalar potentials in (8) depend on the current radius, for which, in view of (9), we can write the following:

$$r = \sqrt{x^2 + y^2 + z^2} = \sqrt{v'^2 t^2 + 2v't\rho \cos\theta \sqrt{1 - v'^2/c^2} + \rho^2\left(1 - \frac{v'^2}{c^2}\cos^2\theta\right)}. \tag{10}$$

The element of the moving volume of the particle in the spherical coordinates (9) is defined by the formula $dV_p = J\, d\rho\, d\theta\, d\varphi$, where $J$ is the Jacobian matrix determinant:

$$J = \frac{\partial(x, y, z)}{\partial(\rho, \theta, \varphi)} = \begin{vmatrix} \frac{\partial x}{\partial \rho} & \frac{\partial x}{\partial \theta} & \frac{\partial x}{\partial \varphi} \\ \frac{\partial y}{\partial \rho} & \frac{\partial y}{\partial \theta} & \frac{\partial y}{\partial \varphi} \\ \frac{\partial z}{\partial \rho} & \frac{\partial z}{\partial \theta} & \frac{\partial z}{\partial \varphi} \end{vmatrix}.$$

Defining $J$, in view of (9), we find the volume element $dV_p = \frac{1}{\gamma'} \rho^2 d\rho \sin\theta d\theta d\varphi$. As a result, in (7) for the integral over the volume of the moving particle we have:

$$\int^n \left( \rho_0 \vartheta_{0p} + \rho_0 \psi_{0p} + \rho_{0q} \varphi_{0p} + \rho_0 \wp_{0p} \right) dV =$$
$$= \frac{1}{\gamma'} \int^n \left( \rho_0 \vartheta_{0p} + \rho_0 \psi_{0p} + \rho_{0q} \varphi_{0p} + \rho_0 \wp_{0p} \right) \rho^2 \sin\theta \, d\rho \, d\theta \, d\varphi. \qquad (11)$$

Due to the relativistic effect of length contraction during motion, the volume of the moving particle becomes the Heaviside ellipsoid. The equation of the surface of such an ellipsoid follows from Lorentz transformations:

$$\frac{(x - v't)^2}{1 - v'^2/c^2} + y^2 + z^2 = a_p^2. \qquad (12)$$

After substituting (9) into (12), it becomes clear that during integration in (11) the coordinate $\rho$ must change from 0 to the particle radius $a_p$, and the angles $\theta$ and $\varphi$ change in the same way as in the spherical coordinates (from 0 to $\pi$ for the angle $\theta$, and from 0 to $2\pi$ for the angle $\varphi$).

Substituting (10) into (8) and (8) into (11), we find at $t = 0$:

$$\int\limits^{n}\left(\rho_{0}\vartheta_{0p}+\rho_{0}\psi_{0p}+\rho_{0q}\varphi_{0p}+\rho_{0}\wp_{0p}\right)dV=$$

$$=\frac{\rho_{0}c^{2}\gamma_{cp}V_{p}}{\gamma'}\left(\begin{array}{l}1-\dfrac{2\pi\eta\rho_{0s}a_{p}^{2}}{5c^{2}}\left(1-\dfrac{v'^{2}}{3c^{2}}\right)+\dfrac{2\pi G\rho_{0s}a_{p}^{2}}{5c^{2}}\left(1-\dfrac{v'^{2}}{3c^{2}}\right)-\dfrac{2\pi G\rho_{0s}a_{p}^{2}}{c^{2}}+\\ +\dfrac{\rho_{0q}\rho_{0qs}a_{p}^{2}}{2\varepsilon_{0}\rho_{0}c^{2}}-\dfrac{\rho_{0q}\rho_{0qs}a_{p}^{2}}{10\varepsilon_{0}\rho_{0}c^{2}}\left(1-\dfrac{v'^{2}}{3c^{2}}\right)+\dfrac{\wp_{cp}}{c^{2}\gamma_{cp}}-\dfrac{2\pi\sigma\rho_{0s}a_{p}^{2}}{5c^{2}}\left(1-\dfrac{v'^{2}}{3c^{2}}\right)\end{array}\right).$$

(13)

Here $V_{p}=\dfrac{4\pi a_{p}^{3}}{3}$ is the invariant volume of the particle. The expression in brackets in (13) depends on the particle radius $a_{p}$, as well as on the ratio $\dfrac{v'^{2}}{3c^{2}}$. This ratio is small in most cases, when the particle's speed $v'$ is significantly less than the speed of light. If we use the notation

$$B=\rho_{0}c^{2}\gamma_{cp}\left(\begin{array}{l}1-\dfrac{2\pi\eta\rho_{0s}a_{p}^{2}}{5c^{2}}\left(1-\dfrac{v'^{2}}{3c^{2}}\right)+\dfrac{2\pi G\rho_{0s}a_{p}^{2}}{5c^{2}}\left(1-\dfrac{v'^{2}}{3c^{2}}\right)-\dfrac{2\pi G\rho_{0s}a_{p}^{2}}{c^{2}}+\\ +\dfrac{\rho_{0q}\rho_{0qs}a_{p}^{2}}{2\varepsilon_{0}\rho_{0}c^{2}}-\dfrac{\rho_{0q}\rho_{0qs}a_{p}^{2}}{10\varepsilon_{0}\rho_{0}c^{2}}\left(1-\dfrac{v'^{2}}{3c^{2}}\right)+\dfrac{\wp_{cp}}{c^{2}\gamma_{cp}}-\dfrac{2\pi\sigma\rho_{0s}a_{p}^{2}}{5c^{2}}\left(1-\dfrac{v'^{2}}{3c^{2}}\right)\end{array}\right).$$

(14)

then (13) can be abbreviated as follows:

$$\int\limits^{n}\left(\rho_{0}\vartheta_{0p}+\rho_{0}\psi_{0p}+\rho_{0q}\varphi_{0p}+\rho_{0}\wp_{0p}\right)dV=\frac{1}{\gamma'}BV_{p}.$$

Substituting this into (7) gives:

$$G_{V}=\frac{1}{c^{2}}\sum_{n=1}^{N}\gamma'\mathbf{r}_{n}\cdot\mathbf{v}'_{n}BV_{p}.$$

Let us turn in this expression from the sum over the particles to the integral over the sphere's volume, assuming that the element of the sphere's volume is the quotient from division of the particle's invariant volume $V_{p}$ by the Lorentz factor $\gamma'$. In the first approximation we can also assume that the quantity $B$ is constant and is the same for all the particles, and then it can be taken outside the integral sign:

$$G_V = \frac{B}{c^2} \int \gamma'^2 \mathbf{r} \cdot \mathbf{v}' dV . \tag{15}$$

Let us take the time derivative of $G_V$ in (15) in such a way that the quantity $\frac{d(\gamma' \mathbf{v}')}{dt}$ would appear:

$$\frac{dG_V}{dt} = \frac{B}{c^2} \int \frac{d}{dt}\left(\gamma'^2 \mathbf{r} \cdot \mathbf{v}'\right) dV = \frac{B}{c^2} \int \gamma' \mathbf{v}' \cdot \frac{d(\gamma' \mathbf{r})}{dt} dV + \frac{B}{c^2} \int \gamma' \mathbf{r} \cdot \frac{d(\gamma' \mathbf{v}')}{dt} dV . \tag{16}$$

### 4. Calculation for relation (16)

As in [5], the time derivative inside the integral on the left-hand side of (16) will be considered as a material derivative:

$$\frac{d\left(\gamma'^2 \mathbf{r} \cdot \mathbf{v}'\right)}{dt} = \frac{\partial\left(\gamma'^2 \mathbf{r} \cdot \mathbf{v}'\right)}{\partial t} + \mathbf{v} \cdot \nabla\left(\gamma'^2 \mathbf{r} \cdot \mathbf{v}'\right) . \tag{17}$$

Since we must take everywhere the average values for typical particles, the relation $\mathbf{v} \cdot \nabla = v_x \frac{\partial}{\partial x} + v_y \frac{\partial}{\partial y} + v_z \frac{\partial}{\partial z}$ should be considered statistically, assuming that the amplitudes of individual terms are equal to each other, that is, $v_x \frac{\partial}{\partial x} = v_y \frac{\partial}{\partial y} = v_z \frac{\partial}{\partial z} = v_r \frac{\partial}{\partial r}$. Then, in the first approximation, we can assume that $\mathbf{v} \cdot \nabla = 3 v_r \frac{\partial}{\partial r}$, where $v_r$ is the averaged radial component of the velocity $\mathbf{v}'$. In this case, all the variables will depend only on the radial coordinate $r$, so that $\frac{\partial\left(\gamma'^2 \mathbf{r} \cdot \mathbf{v}'\right)}{\partial t} = 0$, as well as $\mathbf{r} \cdot \mathbf{v}' = r v_r$. In view of (17), for the left-hand side of (16) in the spherical coordinates, where $dV = r^2 dr\, d\phi \sin\theta d\theta$, we can write:

$$\frac{dG_V}{dt} = \frac{B}{c^2} \int \frac{d}{dt}\left(\gamma'^2 \mathbf{r} \cdot \mathbf{v}'\right) dV = \frac{12\pi B}{c^2} \int v_r \frac{\partial}{\partial r}\left(\gamma'^2 r v_r\right) r^2 dr . \tag{18}$$

The right-hand side of (16) contains the following time derivative:

$$\frac{d(\gamma'\mathbf{r})}{dt} = \gamma'\frac{d\mathbf{r}}{dt} + \mathbf{r}\frac{d\gamma'}{dt} = \gamma'\mathbf{v}' + \mathbf{r}\frac{\partial\gamma'}{\partial t} + \mathbf{r}(\mathbf{v}\cdot\nabla\gamma').$$

Since the Lorentz factor $\gamma' = \dfrac{1}{\sqrt{1-v'^2/c^2}}$ is expressed in terms of the square of the velocity $v'^2$, the relation $\gamma'^2 v'^2 = \gamma'^2 c^2 - c^2$ will hold true. Also taking into account the relations $\dfrac{\partial\gamma'}{\partial t} = 0$, $\mathbf{v}\cdot\nabla = 3v_r\dfrac{\partial}{\partial r}$ and $\mathbf{r}\cdot\mathbf{v}' = rv_r$, for the first integral on the right-hand side of (16) in the spherical coordinates we find:

$$\frac{B}{c^2}\int \gamma'\mathbf{v}'\cdot\frac{d(\gamma'\mathbf{r})}{dt}dV = \frac{4\pi B}{c^2}\int\left(\gamma'^2 c^2 - c^2 + 3\gamma' r v_r^2 \frac{\partial\gamma'}{\partial r}\right) r^2\, dr. \qquad (19)$$

In the system under consideration, due to randomness of the particles' motion, the global vector field potentials and the corresponding solenoidal vectors are close to zero, including the gravitational torsion field $\mathbf{\Omega}$, the magnetic field induction $\mathbf{B}$, and the solenoidal vector of the pressure field $\mathbf{I}$. In this case the Lorentz forces are absent and, similarly to [7], the force densities in the equation of the particles' motion depend only on the field strengths:

$$\frac{d(\gamma'\mathbf{v}')}{dt} = \mathbf{\Gamma} + \frac{\rho_{0q}}{\rho_0}\mathbf{E} + \mathbf{C}. \qquad (20)$$

Equation (20) is valid in the approximation of the special theory of relativity, while the motion of typical particles is approximated by rectilinear motion without proper rotation of the particles.

The strengths of the electric, gravitational and pressure fields inside the sphere, acting on the typical particles, were found in [9]:

$$\boldsymbol{\Gamma} = -\frac{Gc^2 \gamma_c \mathbf{r}}{\eta r^3}\left[\frac{c}{\sqrt{4\pi\eta\rho_0}}\sin\left(\frac{r}{c}\sqrt{4\pi\eta\rho_0}\right) - r\cos\left(\frac{r}{c}\sqrt{4\pi\eta\rho_0}\right)\right] \approx$$

$$\approx -\frac{4\pi G \rho_0 \gamma_c \mathbf{r}}{3}\left(1 - \frac{4\pi\eta\rho_0 r^2}{10c^2}\right).$$

$$\mathbf{E} = \frac{\rho_{0q} c^2 \gamma_c \mathbf{r}}{4\pi\varepsilon_0 \eta \rho_0 r^3}\left[\frac{c}{\sqrt{4\pi\eta\rho_0}}\sin\left(\frac{r}{c}\sqrt{4\pi\eta\rho_0}\right) - r\cos\left(\frac{r}{c}\sqrt{4\pi\eta\rho_0}\right)\right] \approx$$

$$\approx \frac{\rho_{0q} \gamma_c \mathbf{r}}{3\varepsilon_0}\left(1 - \frac{4\pi\eta\rho_0 r^2}{10c^2}\right).$$

$$\mathbf{C} = \frac{\sigma c^2 \gamma_c \mathbf{r}}{\eta r^3}\left[\frac{c}{\sqrt{4\pi\eta\rho_0}}\sin\left(\frac{r}{c}\sqrt{4\pi\eta\rho_0}\right) - r\cos\left(\frac{r}{c}\sqrt{4\pi\eta\rho_0}\right)\right] \approx$$

$$\approx \frac{4\pi\sigma\rho_0 \gamma_c \mathbf{r}}{3}\left(1 - \frac{4\pi\eta\rho_0 r^2}{10c^2}\right).$$

Let us substitute these expressions for the field strengths into (20):

$$\frac{d(\gamma' \mathbf{v})}{dt} = -\frac{c^2 \gamma_c \mathbf{r}}{r^3}\left[\frac{c}{\sqrt{4\pi\eta\rho_0}}\sin\left(\frac{r}{c}\sqrt{4\pi\eta\rho_0}\right) - r\cos\left(\frac{r}{c}\sqrt{4\pi\eta\rho_0}\right)\right]. \tag{21}$$

In derivation of (21), we used the relation between the field coefficients obtained in [12] with the help of the equation of motion and the generalized Poynting theorem:

$$\eta + \sigma = G - \frac{\rho_{0q}^2}{4\pi\varepsilon_0 \rho_0^2}. \tag{22}$$

Substitution of (21) into the second integral on the right-hand side of (16) in the spherical coordinates gives the following:

$$\frac{B}{c^2}\int \gamma' \mathbf{r} \cdot \frac{d(\gamma' \mathbf{v}')}{dt} dV = -4\pi B \gamma_c \int \frac{\gamma'}{r}\left[\frac{c}{\sqrt{4\pi\eta\rho_0}}\sin\left(\frac{r}{c}\sqrt{4\pi\eta\rho_0}\right) - r\cos\left(\frac{r}{c}\sqrt{4\pi\eta\rho_0}\right)\right] r^2 dr.$$

(23)

## 5. The radial velocity component

We will substitute (18), (19) and (23) into (16), cancel out the identical factors, remove the integrals, and after subtracting the identical terms we will obtain:

$$3\gamma' v_r \frac{\partial}{\partial r}(\gamma' r v_r) = \frac{3}{2r}\frac{\partial}{\partial r}(\gamma' r v_r)^2 =$$

$$= \gamma'^2 c^2 - c^2 - \frac{c^2 \gamma_c \gamma'}{r}\left[\frac{c}{\sqrt{4\pi\eta\rho_0}}\sin\left(\frac{r}{c}\sqrt{4\pi\eta\rho_0}\right) - r\cos\left(\frac{r}{c}\sqrt{4\pi\eta\rho_0}\right)\right].$$

This equality represents a differential equation, which allows us to find $v_r$. We will substitute $\gamma'$ from (5) into the right-hand side of the equality:

$$\frac{\partial}{\partial r}(\gamma' r v_r)^2 = \frac{c^3 \gamma_c^2}{3\sqrt{4\pi\eta\rho_0}}\sin\left(\frac{2r}{c}\sqrt{4\pi\eta\rho_0}\right) - \frac{2c^2 r}{3}.$$

Solving this equation, in view of (5), we find its solution:

$$v_r = \frac{c}{\sqrt{3}}\sqrt{1 - \frac{4\pi\eta\rho_0 r^2}{c^2 \gamma_c^2 \sin^2\left(\frac{r}{c}\sqrt{4\pi\eta\rho_0}\right)}} \approx \frac{v_c}{\sqrt{3}}\left(1 - \frac{2\pi\eta\rho_0 r^2}{3v_c^2}\right). \quad (24)$$

An approximate solution on the right-hand side of (24) is obtained by expanding the sine to the second-order terms and taking into account the equality $\gamma_c = \frac{1}{\sqrt{1 - v_c^2/c^2}} \approx 1 + \frac{v_c^2}{2c^2}$ for the Lorentz factor and the speed $v_c$ of the particles at the center of the sphere.

Earlier, we have already estimated the radial velocity component in [5]:

$$v_r \approx \frac{v_c}{\sqrt{3}}\left(1 - \frac{2\pi\eta\rho_0 r^2}{3v_c^2}\right). \quad (25)$$

The coefficient $\sqrt{3}$ in (25) is a consequence of the assumption that, on the average, all the mutually orthogonal components of the velocity of a typical particle in the spherical coordinates have the same amplitude, that is, $v_r \approx v_\theta \approx v_\varphi$. From comparison of (24) and (25) it follows that in (24) the formula for the radial velocity component $v_r$ of the particles was obtained with higher accuracy.

Taking into account that $\gamma' = \dfrac{1}{\sqrt{1 - v'^2/c^2}}$, $v' = c\sqrt{1 - \dfrac{1}{\gamma'^2}}$, with the help of (5) we can find the root-mean-square speed $v'$ of the typical particles inside the sphere:

$$v' = c\sqrt{1 - \frac{4\pi\eta\rho_0 r^2}{c^2 \gamma_c^2 \sin^2\left(\dfrac{r}{c}\sqrt{4\pi\eta\rho_0}\right)}} . \tag{26}$$

Comparing $v'$ and $v_r$ in (24), we find that $v'^2 = 3v_r^2 = v_r^2 + v_\theta^2 + v_\varphi^2$, which exactly corresponds to the assumption of equality of the velocity components' amplitudes $v_r \approx v_\theta \approx v_\varphi$. It should be noted that for an ideal gas the standard method to determine the root-mean-square speed $v_{rms}$ of the particles is to use the Maxwell distribution, when the speed is related to the particles' mass $m$, the temperature $T$, and the Boltzmann constant $k$:

$$v_{rms} = \sqrt{\frac{3kT}{m}} . \tag{27}$$

If we equate (26) and (27), then we can see that at the center of the system the temperature must be maximal.

In the kinetic theory of gases, the speed of sound $v_s$ is expressed in terms of the heat capacity ratio $\delta = \dfrac{c_p}{c_v} = 1 + \dfrac{2}{f}$, where $c_p$ and $c_v$ denote the specific heat capacities at constant pressure and constant volume, respectively; $f$ is the number of degrees of freedom of a gas particle [13]:

$$v_s = v_{rms}\sqrt{\frac{\delta}{3}} = v_{rms}\sqrt{\frac{f+2}{3f}} . \tag{28}$$

With the help of the speed of sound $v_s$ in (28) it is possible to experimentally estimate the root-mean-square speed $v_{rms}$ at different points of the system and compare it with (27). In turn, comparison of $v_{rms}$ with the speed $v'$ in (26) allows us to estimate the acceleration field coefficient $\eta$.

### 6. The equation of motion

With the help of the radial component of the velocity of typical particles (24), we can check if equation of motion (20) is satisfied. Let us project this equation on the radial direction. According to (21), the projection of the right-hand side of (20) is equal to:

$$\Gamma_r + \frac{\rho_{0q}}{\rho_0} E_r + C_r = -\frac{c^2 \gamma_c}{r^2}\left[\frac{c}{\sqrt{4\pi\eta\rho_0}}\sin\left(\frac{r}{c}\sqrt{4\pi\eta\rho_0}\right) - r\cos\left(\frac{r}{c}\sqrt{4\pi\eta\rho_0}\right)\right]. \quad (29)$$

The projection of the left-hand side of (20) on the radial direction, taking into account the transformation $\mathbf{v}\cdot\nabla = 3v_r\frac{\partial}{\partial r}$, used in derivation of (18), and in view of (5) and (24) takes the following form:

$$\frac{d(\gamma' v_r)}{dt} = \frac{\partial(\gamma' v_r)}{\partial t} + \mathbf{v}\cdot\nabla(\gamma' v_r) = 3v_r \frac{\partial(\gamma' v_r)}{\partial r} =$$

$$= \frac{c^2 \gamma_c}{r}\cos\left(\frac{r}{c}\sqrt{4\pi\eta\rho_0}\right) - \frac{c^3 \gamma_c}{r^2\sqrt{4\pi\eta\rho_0}}\sin\left(\frac{r}{c}\sqrt{4\pi\eta\rho_0}\right).$$

From the equality of the given expression and (29) it follows that (24) agrees with the equation of motion.

### 7. Quantitative relations in the generalized virial theorem

Similarly to (2), we will take the time derivative of the quantity $G_V$ in (15):

$$\frac{dG_V}{dt} = \frac{B}{c^2}\int \frac{d}{dt}\left(\gamma'^2 \mathbf{r}\cdot\mathbf{v}'\right)dV = \frac{B}{c^2}\int \mathbf{r}\cdot\frac{d(\gamma'^2 \mathbf{v}')}{dt}dV + \frac{B}{c^2}\int \gamma'^2 \mathbf{v}'\cdot\frac{d\mathbf{r}}{dt}dV. \quad (30)$$

Let us calculate each term in (30) and then compare these terms with each other. For the left-hand side of (30), in view of (18), (24) and (5), we find:

$$\frac{dG_V}{dt} = \frac{12\pi B}{c^2} \int_0^a v_r \frac{\partial}{\partial r}\left(\gamma'^2 r v_r\right) r^2 dr =$$

$$= \frac{Bc^2 \gamma_c^2}{2\eta \rho_0} \left[ \frac{c\sin\left(\frac{2a}{c}\sqrt{4\pi\eta\rho_0}\right)}{\sqrt{4\pi\eta\rho_0}} - a - a\cos\left(\frac{2a}{c}\sqrt{4\pi\eta\rho_0}\right) \right] - \quad (31)$$

$$- \frac{4\pi B\sqrt{4\pi\eta\rho_0}}{c} \int_0^a r^3 \operatorname{ctg}\left(\frac{r}{c}\sqrt{4\pi\eta\rho_0}\right) dr.$$

Let us transform the expression under integral sign on the right-hand side of (30) with the help of (21) and the equalities $\mathbf{r}\cdot\mathbf{v}' = r v_r$ and $\mathbf{v}\cdot\nabla = 3 v_r \frac{\partial}{\partial r}$:

$$\mathbf{r}\cdot\frac{d(\gamma'^2 \mathbf{v}')}{dt} = \gamma' \mathbf{r}\cdot\frac{d(\gamma' \mathbf{v}')}{dt} + \mathbf{v}'\cdot\mathbf{r}\gamma'\frac{d\gamma'}{dt} = \gamma' \mathbf{r}\cdot\frac{d(\gamma' \mathbf{v}')}{dt} + v_r r\gamma'(\mathbf{v}\cdot\nabla)\gamma' =$$

$$= -\frac{c^2 \gamma_c \gamma'}{r}\left[\frac{c}{\sqrt{4\pi\eta\rho_0}}\sin\left(\frac{r}{c}\sqrt{4\pi\eta\rho_0}\right) - r\cos\left(\frac{r}{c}\sqrt{4\pi\eta\rho_0}\right)\right] + 3 v_r^2 r \gamma' \frac{\partial \gamma'}{\partial r}.$$

Taking into account this relation and relations (5) and (24), for the first integral on the right-hand side of (30) in the spherical coordinates we obtain:

$$\frac{B}{c^2}\int \mathbf{r}\cdot\frac{d(\gamma'^2 \mathbf{v}')}{dt} dV =$$

$$= \frac{4\pi B}{c^2}\int_0^a \left\{ -\frac{c^2 \gamma_c \gamma'}{r}\left[\frac{c}{\sqrt{4\pi\eta\rho_0}}\sin\left(\frac{r}{c}\sqrt{4\pi\eta\rho_0}\right) - r\cos\left(\frac{r}{c}\sqrt{4\pi\eta\rho_0}\right)\right] + 3 v_r^2 r \gamma' \frac{\partial \gamma'}{\partial r} \right\} r^2 dr =$$

$$= \frac{Bc^2 \gamma_c^2}{2\eta \rho_0}\left[ \frac{3c\sin\left(\frac{2a}{c}\sqrt{4\pi\eta\rho_0}\right)}{2\sqrt{4\pi\eta\rho_0}} - a\cos\left(\frac{2a}{c}\sqrt{4\pi\eta\rho_0}\right) - 2a + \frac{8\pi\eta\rho_0 a^3}{3c^2 \gamma_c^2} \right] -$$

$$- \frac{4\pi B\sqrt{4\pi\eta\rho_0}}{c}\int_0^a \left[ r^3 \operatorname{ctg}\left(\frac{r}{c}\sqrt{4\pi\eta\rho_0}\right) \right] dr.$$

(32)

We will calculate the last integral on the right-hand side of (30) with the help of (5) and the relation $\gamma'^2 v'^2 = c^2 \gamma'^2 - c^2$:

$$\frac{B}{c^2}\int \gamma'^2 \mathbf{v}' \cdot \frac{d\mathbf{r}}{dt} dV = \frac{B}{c^2}\int \gamma'^2 v'^2 \, dV = 4\pi B \int (\gamma'^2 - 1) r^2 dr =$$
$$= \frac{Bc^2 \gamma_c^2}{2\eta \rho_0}\left[ a - \frac{c}{2\sqrt{4\pi \eta \rho_0}} \sin\left(\frac{2a}{c}\sqrt{4\pi \eta \rho_0}\right)\right] - \frac{4\pi B a^3}{3}. \tag{33}$$

If we substitute (31), (32) and (33) into (30), then all the terms are canceled out without a remainder, which proves the correctness of our calculations. Now we will expand the periodic functions in (31), (32) and (33) to the terms containing $c^4$ in the denominator. Then we will use the approximate expression for the Lorentz factor in terms of the square of the velocity of the particles at the center of the sphere, according to [5]:

$$\gamma_c^2 = \frac{1}{1 - v_c^2/c^2} \approx 1 + \frac{v_c^2}{c^2} \approx 1 + \frac{3\eta m}{5ac^2}\left(1 + \frac{9}{\sqrt{56}}\right), \tag{34}$$

where $m$ is the product of the mass density $\rho_0$ by the volume of the sphere with the radius $a$.

This gives us the following:

$$\frac{dG_V}{dt} = \frac{12\pi B}{c^2}\int_0^a v_r \frac{\partial}{\partial r}\left(\gamma'^2 r v_r\right) r^2 dr \approx$$
$$\approx \frac{4\pi a^3 B \gamma_c^2}{3}\left(1 - \frac{9\eta m}{5ac^2}\right) - \frac{4\pi a^3 B}{3}\left(1 - \frac{3\eta m}{5ac^2}\right) \approx -\frac{4\pi \eta m a^3 B}{5ac^2}\left(1 - \frac{9}{\sqrt{56}}\right). \tag{35}$$

$$\frac{B}{c^2}\int \mathbf{r} \cdot \frac{d(\gamma'^2 \mathbf{v}')}{dt} dV \approx -\frac{8\pi \eta m a^3 B \gamma_c^2}{5ac^2} + \frac{4\pi \eta m a^3 B}{5ac^2} \approx -\frac{4\pi \eta m a^3 B}{5ac^2}. \tag{36}$$

$$\frac{B}{c^2}\int \gamma'^2 \mathbf{v}' \cdot \frac{d\mathbf{r}}{dt} dV \approx \frac{4\pi a^3 B \gamma_c^2}{3}\left(1 - \frac{3\eta m}{5ac^2}\right) - \frac{4\pi a^3 B}{3} \approx \frac{36\pi \eta m a^3 B}{5\sqrt{56}\, ac^2}. \tag{37}$$

According to (14), $B \approx \rho_0 c^2 \gamma_{cp}$, which gives for (37) the following:

$$\frac{B}{c^2}\int \gamma'^2 \mathbf{v}' \cdot \frac{d\mathbf{r}}{dt} dV = \frac{B}{c^2}\int \gamma'^2 v'^2 \, dV = 4\pi B \int_0^a (\gamma'^2 - 1) r^2 dr \approx \frac{27\eta m^2 \gamma_{cp}}{5\sqrt{56}a}. \quad (38)$$

By its meaning the energy in (37) and (38) corresponds to the energy $\sum_{n=1}^{N}(\mathbf{P}_n \cdot \mathbf{v}_n) = 2E_V$ in (2) and represents the doubled energy of motion $E_V$, as was determined in [14]. In this case, $2E_V$ is equal in its value to the sum of the Lagrangian and the Hamiltonian. Thus

$$E_V \approx \frac{27\eta m^2 \gamma_{cp}}{10\sqrt{56}a}.$$

According to the integral theorem of generalized virial (30), relation (35) is the sum of relations (36) and (37). We see that the time derivative of the virial function $\frac{dG_V}{dt}$ in (35) is not equal to zero. Assuming that energy (36) is the potential energy $W_G$, associated with the generalized forces inside the system, and energy (37) equals $2E_V$, for the relation between these energies we find the following:

$$E_V + \frac{9}{2\sqrt{56}} W_G \approx E_V + 0.6 W_G \approx 0. \quad (39)$$

Relation (39) is the obtained relation between the energies, which differs from the classical virial theorem (3) and coincides exactly with the relation between the energies in [5], where the relativistic virial theorem was studied taking into account four fields. Thus, the generalized virial theorem gives the same result as the relativistic virial theorem.

The kinetic energy of the particles of the system under consideration was found in [5] in the following form:

$$E_k = 4\pi \rho_0 c^2 \int_0^a (\gamma'^2 - \gamma') r^2 dr \approx \frac{mv_c^2 \gamma_c}{2} - \frac{3\eta m^2 \gamma_c}{10a} - \frac{3\eta m^2 v_c^2 \gamma_c}{10ac^2} + \frac{3mv_c^4 \gamma_c}{8c^2}. \quad (40)$$

If we substitute here $v_c^2 \approx \frac{3\eta m}{5a}\left(1 + \frac{9}{\sqrt{56}}\right)$ from (34), the kinetic energy in the first approximation is expressed as follows:

$$E_k \approx \frac{27\eta m^2 \gamma_c}{10\sqrt{56}\, a} \approx E_V.$$

Hence it follows that in the system under consideration the kinetic energy $E_k$ is of the same order as the energy of motion $E_V$. This coincidence will be valid as long as the velocity $\mathbf{v}'$ of the particles remains small in comparison to the speed of light, and the Lorentz factors $\gamma_c$ and $\gamma_{cp}$ are close to unity.

However, in the general case $E_V \neq E_k$, since according to (38) and (40)

$$E_V = 2\pi B \int_0^a (\gamma'^2 - 1)\, r^2 dr \neq 4\pi \rho_0 c^2 \int_0^a (\gamma'^2 - \gamma')\, r^2 dr = E_k. \tag{41}$$

## 8. Conclusion

In contrast to the standard virial theorem, the generalized virial theorem deals with the particles' generalized momenta. We use generalized momenta because the sum of such momenta is conserved in a closed system, as it follows from Lagrange mechanics. In the relativistic uniform system under consideration, the global vector field potentials are equal to zero due to the random motion of the particles. Therefore, the generalized momenta of the particles include only the proper vector potentials of the fields arising from the particles' motion.

Since the time derivative of the virial function (30) turns out to be nonzero, the relation $E_V \approx -0.6 W_G$ in (39) is satisfied instead of (3), and the proportion of the energy of the particles' motion relative to the absolute value of the potential energy, associated with the generalized forces, increases in comparison with the classical case. The results obtained in general coincide with those found in [5] for the ordinary virial theorem in the relativistic form.

We show that the time derivative of the generalized virial function is not equal to zero due to the dependence of this function on the current radius. The time derivative in this case should be considered as a material derivative, including a convective derivative. Besides, as a result of averaging of the quantities in the convective derivative, we can make a substitution of the form $\mathbf{v} \cdot \nabla = 3 v_r \frac{\partial}{\partial r}$, and thus use the radial component of the particles' velocity $v_r$ in the formulation of the theorem. This allows us to find the formula for calculation of $v_r$ in (24) and then to verify it in the equation of motion of the typical particles.

The formula presented in (26) determines the root-mean-square speed $v'$ of typical particles at each point of the system with the help of the current radius and parameters of the particles and is derived using the field theory. This distinguishes significantly $v'$ from the root-mean-square speed $v_{rms}$ (27), expressed in terms of the temperature, and from the speed of sound $v_s$ in (28), found by statistical methods. Since the root-mean-square speed should not depend on the method of its determination, the combination of formulas (26-28) allows us to find new relationships between the particles' parameters and the thermodynamic parameters.

In (38) and (39) we find the energy of motion $E_V$, which is sufficiently close in its magnitude to the kinetic energy $E_k$ in (40). In order to understand the difference between these energies, we will turn to the gauge condition for the relativistic energy according to [7] and [10]:

$$-ck\Lambda = U_\mu J^\mu + D_\mu J^\mu + A_\mu j^\mu + \pi_\mu J^\mu, \qquad (42)$$

where $k = -\dfrac{c^3}{16\pi G\beta}$; $G$ is the gravitational constant; $\beta$ is a constant of the order of unity, which is included in the equation for the metric as a multiplier; $\Lambda$ is the cosmological constant; $U_\mu = \left(\dfrac{\vartheta}{c}, -\mathbf{U}\right)$, $D_\mu = \left(\dfrac{\psi}{c}, -\mathbf{D}\right)$, $A_\mu = \left(\dfrac{\varphi}{c}, -\mathbf{A}\right)$ and $\pi_\mu = \left(\dfrac{\wp}{c}, -\mathbf{\Pi}\right)$ represent the four-potentials for the acceleration field, gravitational field, electromagnetic field and pressure field, respectively; $\vartheta$, $\psi$, $\varphi$ and $\wp$ are the scalar potentials of the acceleration field, gravitational field, electromagnetic field and pressure field, respectively.

When condition (42) is met, the relativistic energy of the system does not depend either on the scalar curvature or on the cosmological constant, and becomes uniquely determined. In (42) the cosmological constant is expressed in terms of the sum of the products of the fields' four-potentials by the four-currents, while all the fields acting in the system are assumed to be vector fields. In particular, the gravitational field is considered in the framework of the vector covariant theory of gravitation, which, in the limit of the weak field and low velocities, is transformed into the Lorentz-invariant theory of gravitation, which generalizes the Newton's theory of gravitation to inertial reference frames. We should note that in the general theory of relativity the gravitational field is a tensor field associated with the metric tensor. This leads to impossibility of using the cosmological constant for the energy gauging similarly to (42), since the interpretation of the cosmological constant itself changes. In this case, it becomes impossible either to localize uniquely the gravitational energy in space, or to calculate the system's energy irrespectively of the choice of the reference frame [15-17].

Within the framework of the special theory of relativity, the mass four-current $J^\mu = \rho_0 u^\mu$, the charge four-current $j^\mu = \rho_{0q} u^\mu$, where $u^\mu = (c\gamma, \mathbf{v}\gamma)$ is the four-velocity. The gauge condition for the energy will have the following form:

$$-ck\Lambda = \gamma\rho_0(\vartheta - \mathbf{v}\cdot\mathbf{U}) + \gamma\rho_0(\psi - \mathbf{v}\cdot\mathbf{D}) + \gamma\rho_{0q}(\varphi - \mathbf{v}\cdot\mathbf{A}) + \gamma\rho_0(\wp - \mathbf{v}\cdot\mathbf{\Pi}). \qquad (43)$$

Let us suppose that the considered spherical system of particles and fields was formed from the matter, which was initially scattered at infinity and was almost motionless there, and then it was collected into a sphere under the action of gravitation. In the initial state, we can assume that the particles' velocities $\mathbf{v} = 0$, then the Lorentz factor for all the particles $\gamma = 1$. In this case the scalar potentials of the fields would be equal to the proper potentials of the particles, and for (43) we have the following:

$$-ck\Lambda = \rho_0 \vartheta_{0p} + \rho_0 \psi_{0p} + \rho_{0q} \varphi_{0p} + \rho_0 \wp_{0p}. \qquad (44)$$

Now we will note that the sum of the terms on the right-hand side of (44) is present in relations (7), (11), (14) as a multiplier, and in all the relations where the quantity $B$ is present. Thus, the cosmological constant $\Lambda$ of the physical system under consideration becomes included in our formulation of the integral theorem of generalized virial.

Let us now consider the fundamental difference between the kinetic energy $E_k$ of the particles and the energy of motion $E_V$, the doubled value of which is provided in (38) according to [7]. The fact is that together with the moving particles, their proper fields associated with the particles are moving as well. These fields have the mass-energy and, consequently, they participate in the formation of the generalized momentum $\mathbf{P}_n$ of each particle. If $E_{kn}$ is the kinetic energy of one particle, then $E_k = \sum_{n=1}^{N} E_{kn}$, however the energy of motion $E_V$ also contains additional contributions from the mass-energy of the particles' fields and therefore it is not equal to the kinetic energy $E_k$. We can also say that the kinetic energy takes into account only the energy of the acceleration field, describing the motion of the particles, and addition of the contribution from all the other fields leads to the energy of motion $E_V$ of the particles and fields of the system. Indeed, $E_k$ in (40) depends on the Lorentz factor $\gamma'$, found according to (5) with the help of the equations for the acceleration field. At the same time, $E_v$ depends on all the fields, since it is determined using the half-sum of the system's Lagrangian and Hamiltonian, and, in addition, it depends on the quantity $B$ and the terms in (44).

The inequality between the energies $E_k$ and $E_v$ in (41) is determined by the system's parameters, however, there is a correlation between these energies, because relation (22) for the fields' coefficients holds true in the system under consideration, and the kinetic and potential energies can be converted into one another according to [18].